\documentclass[]{aa}

\usepackage{natbib}
\bibpunct{(}{)}{;}{a}{}{,} 
\usepackage{graphicx}
\usepackage{txfonts}
\begin{document}
   \title{Clouds in the atmospheres of extrasolar planets}

   \subtitle{III. Impact of low and high-level clouds on the reflection spectra of\\ Earth-like planets}

   \author{D. Kitzmann
          \inst{1},
          A.B.C. Patzer
          \inst{1},
          P. von Paris
          \inst{2},
          M. Godolt
          \inst{1},
          \and
          H. Rauer
          \inst{1,2}
          }
          
   \authorrunning{D. Kitzmann et al.}
   \titlerunning{Clouds in the atmospheres of extrasolar planets. III.}


   \institute{Zentrum f\"ur Astronomie und Astrophysik, Technische Universit\"at Berlin,
              Hardenbergstr. 36, 10623 Berlin (Germany)\\
              \email{kitzmann@astro.physik.tu-berlin.de}
              \and
              Institut f\"ur Planetenforschung, Deutsches Zentrum f\"ur Luft- und Raumfahrt (DLR),
              Rutherfordstr. 2, 12489 Berlin (Germany)
             }

   \date{Received 30 May 2011 / Accepted 2 August 2011}

 
  \abstract
   {Owing to their wavelength dependent absorption and scattering properties, clouds have an important influence on spectral albedos and planetary reflection spectra. In addition, the spectral energy distribution of the incident stellar light determines the detectable absorption bands of atmospheric molecules in these reflection spectra.}
   {We study the influence of low-level water and high-level ice clouds on low-resolution reflection spectra and planetary albedos of Earth-like planets orbiting different types of stars in both the visible and near infrared wavelength range.}
   {We use a one-dimensional radiative-convective steady-state atmospheric model coupled with a parametric cloud model, based on observations in the Earth's atmosphere to study the effect of both cloud types on the reflection spectra and albedos of Earth-like extrasolar planets at low resolution for various types of central stars.}
   {We find that the high scattering efficiency of clouds substantially causes both the amount of reflected light and the related depths of the absorption bands to be substantially larger than in comparison to the respective clear sky conditions. Low-level clouds have a stronger impact on the spectra than the high-level clouds because of their much larger scattering optical depth. The detectability of molecular features in near the UV - near IR wavelength range is strongly enhanced by the presence of clouds. However, the detectability of various chemical species in low-resolution reflection spectra depends strongly on the spectral energy distribution of the incident stellar radiation. In contrast to the reflection spectra the spectral planetary albedos enable molecular features to be detected without a direct influence of the spectral energy distribution of the stellar radiation. Here, clouds increase the contrast between the radiation fluxes of the planets and the respective central star by about one order of magnitude, but the resulting contrast values are still too low to be observable with the current generation of telescopes.}
   {}


   \maketitle
%

\section{Introduction}

The energy budget of terrestrial planetary atmospheres is determined by the balance between the radiative losses of thermal radiation from the planetary surface and atmosphere to space and the absorbed incident stellar radiation.
Clouds have a large impact on this energy budget, hence on the planetary climate and habitability (i.e. the possible existence of liquid water on the planetary surface). They reflect incident radiation from the central star back to space (albedo effect), but also reduce losses of thermal radiation to space (greenhouse effect). As a direct consequence, clouds also have a large impact on the planetary spectra. 
They diminish the thermal IR emission from the planet, while also affecting the reflected incident stellar radiation at near UV (NUV) to near IR (NIR) wavelengths of the planetary spectrum by increasing the amount of back-scattered stellar radiation. 

Our planet Earth provides a unique opportunity to explore the detectability of spectral signatures of atmospheric chemical species in the presence of clouds. Observations of the spectral albedos and reflection spectra of Earth in the visible (VIS) and NIR wavelength range are often done via earthshine observations, i.e. by analysing the spectrum of Earth reflected by the lunar surface \citep[see e.g.][]{Palle2005ApJ,Woolf2002ApJ,Hamdani2005conf}. In principle, signals of surface vegetation ("vegetation red edge") are present in the spectral albedo of Earth. However, as pointed out by for instance \citet{Arnold2002A&A} and \citet{Hamdani2006A&A}, this spectral feature can easily be concealed by clouds. \citet{Palle2006ApJ} and \citet{Tinetti2006ApJ} investigated the detectability of vegetation signatures of extrasolar planets, concluding that clouds play a crucial role for these signatures in the reflection spectra.
Thus, apart from the scattering characteristics of different planetary surface types, the presence of clouds has been found to be one of the most important factors controlling the form of the reflection spectra.

Depending on the type and amount of clouds, reflection spectra allow for the detection of several important atmospheric chemical species \citep{Tinetti2006a, Turnbull2006ApJ}. Corresponding calculations of synthetic high-resolution reflectance spectra of Earth including the effects of different cloud types have been performed by e.g. \citet{Tinetti2006b} and compared to satellite measurements of the reflection spectrum of Earth. Different surface characteristics, solar zenith angles, satellite viewing angles, cloud types, and measured atmospheric temperature profiles were used by \citet{Tinetti2006b} to calculate the synthetic reflection spectra over single points of the Earth surface with a line-by-line radiative transfer code (SMART, see \citet{Meadows1996JGR} and \citet{Crisp1997GeoRL}). These single spectra were then disk-integrated using the observational setup of the satellite measurements (Earth illumination by the Sun, satellite viewing angle, etc.) to obtain the disk-averaged reflection spectrum as seen by the satellite. These observations and synthetic spectra show that clouds can strongly increase the spectral albedo in the visible wavelength and also considerably alter the shape of the spectrum compared to the clear-sky case. The spectral features of molecules detectable in a low-resolution reflection spectrum of a cloud-free habitable, terrestrial extrasolar planet were discussed by \citet{Kaltenegger2009}. Approximations for the effects of clouds were also included in the calculated spectra for different evolutionary stages of the Earth atmosphere by \citet{Kaltenegger2007ApJ}. However, multiple scattering by cloud particles was not considered. Spectra for different present-day Earth clouds were instead added to the clear-sky radiative transfer calculations. The atmospheric profiles used to calculate the spectra were taken from (cloud-free) atmospheric model calculations published by various authors (see \citet{Kaltenegger2007ApJ} for details).

Investigating the polarised radiation scattered by clouds in contrast to the unpolarised stellar radiation provides another opportunity to characterise cloudy exoplanetary atmospheres. The effects of clouds on these polarisation spectra of extrasolar planets were studied among others by \citet{Stam2008A&A} and \citet{Karalidi2011}, who demonstrated that cloud properties (e.g. shape, size) can be obtained from polarisation spectra.

To complement these studies of the atmospheres of terrestrial planets, the effects of clouds on thermal emission, transmission, and reflection spectra have been studied in the atmospheres of brown dwarfs or giant exoplanets (see e.g. \citet{deKok2011A&A}, \citet{Burrows2008ApJ}, \citet{Marley1999ApJ}, \citet{Sudarsky2000ApJ}, or \citet{Helling2008A&A} for details). However, these kinds of atmospheres conceptually differ from those of terrestrial planets in terms of e.g. atmospheric composition and structure, boundary conditions (e.g. existence of a planetary surface), chemistry, and the properties of the condensating species, such that a direct comparison with these atmospheres is difficult. Nonetheless, effective cloud formation in these atmospheres can, for example, also significantly alter the thermal emission and reflection spectra of these objects by means of both the scattering and absorption of radiation caused by cloud particles.

In this paper, we study the influence of low- and high-level clouds on the reflectance spectra and spectral albedos of Earth-like planets in the visible and NIR wavelength range at low resolution. A one-dimensional (1D) steady-state radiative-convective atmospheric model is used in the model calculations. The model accounts for two different cloud layers (low-level water droplet and high-level ice particle clouds) and for the partial overlap of these two layers. A more detailed model description is given in Sect. \ref{sec_model}. To verify the applicability of our modelling approach, this coupled cloud-atmosphere model is applied to the modern Earth atmosphere and its spectral appearance, as described in Sect. \ref{sec_earth_ref}. 
The resulting reflection spectra and spectral albedos of Earth-like planets orbiting different types of central stars and their implications for the detectability of characteristic molecular signatures are presented in Sect. \ref{sec_spectra}.

\section{Details of the model}
\label{sec_model}

The impact of clouds on the thermal IR emission spectra was investigated in detail by \citet{Kitzmann2011A&A531} (henceforth called Paper II) for Earth--like planets around different types of stars. However, clouds also affect the reflected incident stellar radiation in the short wavelength range, from the NUV to NIR, of the planetary spectrum. We study here the influence of clouds on planetary reflection spectra at low resolution using a 1D steady-state radiative-convective atmospheric model developed to account for the radiative effects of multi-layered clouds and their impact on the surface temperature in atmospheres of Earth-like planets \citep[see][Paper I, for details]{Kitzmann2010}.

We adopt a parameterised description of two different cloud layers, low-level droplet and high-level ice particle clouds, that is derived from in-situ measurements of the respective cloud type in the atmosphere of Earth. This cloud model is included into the originally cloud-free atmospheric model of \citet{Kasting1984}, in its form developed by \citet{Pavlov00} and \citet{Segura03}. To limit the number of cloud parameters, the minimal possible partial overlap of both cloud layers is assumed in the calculations unless otherwise stated. The altitude of each cloud layer is not simply fixed in height, but iteratively adjusted to match the corresponding measured Earth pressure values (low-level water cloud: $0.83~\mathrm{bar}$, high-level ice cloud: $0.27~\mathrm{bar}$). The freezing limit of the water droplets and the limiting temperature for liquefying the ice particles determine the range of temperatures for which our method is valid (cf. Paper I)\footnote{Note that other atmospheric feedback effects on the clouds are not explicitly taken into account.}. For all calculations, the same chemical composition of the atmosphere is assumed, which is chosen to represent modern Earth conditions \citep[see][]{Grenfell07}.
\begin{figure}
  \centering
  \resizebox{\hsize}{!}{\includegraphics{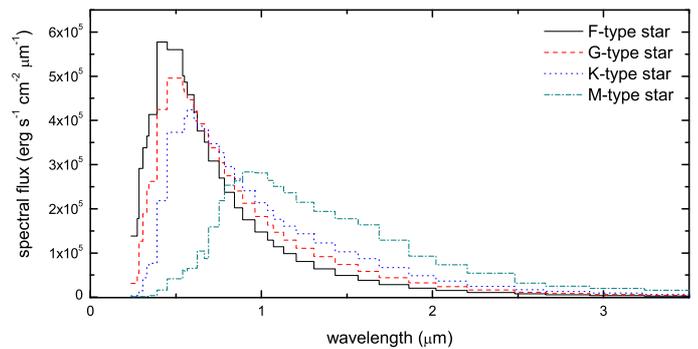}}
  \caption{Incident stellar spectra for the different central stars. Each radiation flux is scaled to the distance where the total stellar energy input at the top of the planetary atmosphere equals the solar constant. The high-resolution spectra have been obtained from measurements in the UV and complemented at longer wavelengths with synthetic spectra. The high-resolution spectra have been binned according to the spectral resolution of the radiative transfer used in the atmospheric model (see text).}
  \label{stellar_spectra}
\end{figure}
The treatment of the radiation transfer problem in the atmospheric model is optimised for the energy transport in Earth-like atmospheres \citep[e.g.][]{Mlawer97}. As usual, the calculation of the radiation transport is divided into two spectral parts. The first part treats the absorption and scattering of incident solar radiation in the short wavelength regime, the second handles the absorption and emission of thermal radiation from the planetary surface and atmosphere, including the multiple scattering caused by cloud particles. The plane-parallel radiative transfer equation in the short wavelength regime is solved by applying a $\delta$-two-stream quadrature approach \citep{Toon89}, which uses 38 broad spectral intervals between 0.238~$\mu$m and 4.55~$\mu$m with variable spectral resolution ($10 \leq$R$\leq 100$). These intervals define the spectral resolution of the planetary spectra presented here, which are obtained directly from the radiative transfer calculations of the atmospheric model. The molecular absorption of H$_2$O, CO$_2$, O$_2$, O$_3$, and CH$_4$ is treated with four-term correlated-$k$ coefficients \citep[cf.][]{Segura03}. Rayleigh scattering is considered for O$_2$, N$_2$, and CO$_2$ in this wavelength region. A description of the IR radiation transport treatment and the resulting thermal emission spectra is given in Paper II. In addition to the gas opacities, the frequency-dependent optical properties of the cloud particles were previously introduced into the radiative transfer schemes including multiple scattering (Paper I).

To account for different coverages of multi-layered clouds and their partial overlap (being non uni-dimensional quantities) in a 1D atmospheric model, we use a flux-averaging procedure, where the radiative transfer problem is solved for every distinct cloud configuration separately. By averaging these radiation fluxes weighted with the respective cloud cover values, the mean radiative flux is obtained, which then enters into the atmospheric model calculations (see Paper I for details).

Four different central star types are considered in the model calculations: F2V, G2V, K2V, and M4.5V-type stars. The sample of stars used for the calculations are the F-dwarf $\sigma$ Bootis (\object{HD 128167}), the Sun, the young active K-type star $\epsilon \ \mathrm{Eridani}$ (\object{HD 22049}), and the M-type dwarf star AD Leo (\object{GJ 388}). The spectra for the F, K, and M-type stars were obtained from measurements in the UV by the IUE satellite and were complemented at longer wavelengths with synthetic spectra (see Paper I for details). In addition, the optical spectrum from \citet{Pettersen1989} between $335.5 \ \mathrm{nm}$ and $900 \ \mathrm{nm}$ and the NIR spectra of \citet{Leggett1996} are used for the M-star. For the Sun, a measured high-resolution spectrum is used \citep{Gueymard2004}. We note that in general synthetic spectra of different stellar models may display considerable differences in the spectral lines (see e.g. \citet{Sinclair2010MNRAS} for an intercomparison of stellar model spectra). The incident stellar fluxes are scaled by varying the orbital distances, such that the energy integrated over each incident stellar spectrum equals the present Total Solar Irradiance (TSI) at the top of the atmospheres. The corresponding stellar spectra were binned to the spectral intervals of the radiative transfer and are shown in Fig. \ref{stellar_spectra}. According to Paper I, the orbital distances of the Earth-like planets are $1.89 \ \mathrm{AU}$ (F-type star), $1 \ \mathrm{AU}$ (G-type star), $0.61 \ \mathrm{AU}$ (K-type star), and $0.15 \ \mathrm{AU}$ (M-type star).

In our model, the measured terrestrial value of the global mean surface albedo (0.13) is used. A tuned surface albedo is sometimes used to mimic the effect of clouds on the surface temperature in cloud-free model calculations of Earth-like planetary atmospheres \citep[see e.g.][]{Segura03, Segura05}. This adjustment is only feasible if the value of the resulting surface temperature is prescribed as, for example, the measured global mean surface temperature of Earth ($288 \ \mathrm K$). In contrast to this tuning approach, the planetary surface temperature is a result of the calculations in the case of an atmospheric model including clouds.

\section{Reflection spectrum of Earth influenced by high and low-level clouds}
\label{sec_earth_ref}

\begin{figure}
  \centering
  \resizebox{\hsize}{!}{\includegraphics{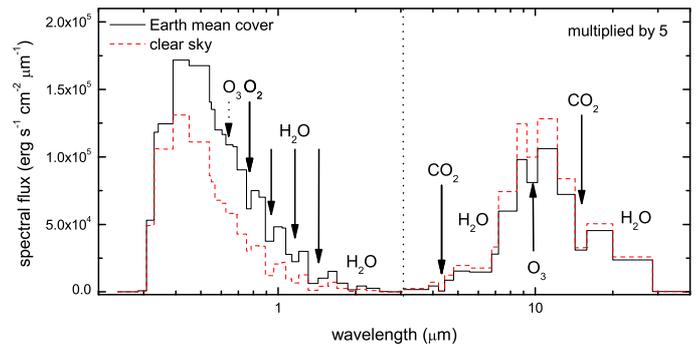}}
  \caption{Planetary spectra at the top of the atmosphere of the Earth model from near UV to IR wavelengths. The dashed line marks the clear sky case, the solid line indicates the results for a mean Earth cloud cover ($39.5\%$ low-level clouds, $15\%$ high-level clouds, $7\%$ overlap of both cloud layers).}
  \label{earth_ref_spectrum}
\end{figure}

To assess the applicability of our model for the calculation of reflection spectra of Earth-like planets around different types of central stars, we study at first the resulting spectra obtained for the Earth model introduced in Paper I. With the observed mean cloud coverages ($39.5\%$ low-level clouds, $15\%$ high-level clouds, and a $7\%$ overlap of both cloud layers), the measured global mean surface temperature of Earth ($288 \ \mathrm{K}$) is reproduced by our model. The clear sky calculation results in a surface temperature of $293 \ \mathrm{K}$, which is clearly too high (see Paper I for a detailed discussion of the effects of clouds on the surface temperature). 

Mid-level clouds are omitted in the cloud model as most of them have been reported to be radiatively neutral in the Earth atmosphere, i.e. their greenhouse and albedo effect balance each other (see \citet{Poetzsch95} and Paper I\&II for details). However, the neglect of the mid-level clouds yields less back-scattered shortwave and more outgoing longwave radiation at the top of the atmosphere than the global energy budget of Earth including all cloud types. In this study, we focus on the effect of high and low-level clouds on the reflection spectra, corresponding to the model introduced in Paper I. These cloud types represent the extreme cases for the cloud effects on the reflection spectra. The observations of e.g. \citet{Tinetti2006a,Tinetti2006b} have shown that the effects of mid-level clouds on the reflection spectra of Earth are between these two extremes. Therefore, the important range of cloud effects on the reflection spectra are covered in our study of extrasolar planetary atmospheres.

The corresponding spectra covering the wavelength range from the near UV to the IR are shown in Fig. \ref{earth_ref_spectrum}. For a better visualisation, the IR part of the spectrum has been multiplied by a factor of 5.
In the visible range ($< 4 \ \mathrm{\mu m}$) the spectrum is a reflection spectrum due to back-scattered, and partly absorbed, incident solar radiation. The emission spectrum in the IR wavelength range originates from the thermal emissions of the planetary surface and the atmosphere, affected by absorption and also scattering in the presence of clouds (see Paper II for details).

The number of chemical species visible in a low-resolution spectrum is rather small. In the IR range, $\mathrm{CO_2}$ ($\sim 15 \ \mathrm{\mu m}$), $\mathrm{H_2 O}$ (vibrational and rotational at $5-8 \ \mathrm{\mu m}$ and $> 15 \ \mathrm{\mu m}$), and the $9.6 \ \mathrm{\mu m}$ $\mathrm{O_3}$ band can be seen at low resolution. The NUV-NIR range allows for the detection of $\mathrm{O_2}$, $\mathrm{H_2 O}$, and in principle also $\mathrm{O_3}$ (see Fig. \ref{earth_ref_spectrum}). Several $\mathrm{H_2 O}$ bands can be found in the reflection spectrum at for instance $\sim 0.95 \ \mathrm{\mu m}$, $\sim 1.1 \ \mathrm{\mu m}$, and $\sim 1.4 \ \mathrm{\mu m}$. The small Fraunhofer A-band of $\mathrm{O_2}$ is visible at $0.76 \ \mathrm{\mu m}$. In principle, the broad Chappuis band of $\mathrm{O_3}$ ($0.55-0.70 \ \mathrm{\mu m}$) is also present, which, however, cannot be directly seen in the calculated low-resolution reflection spectrum of Earth (see Fig. 2 of Paper I for the atmospheric profiles of chemical species). While for instance a spectral resolution of $R=20$ is sufficient to detect the $\mathrm{O_3}$ absorption band at  $9.6 \ \mathrm{\mu m}$ in the IR \citep[see][]{Kaltenegger2009}, it has to be $R>50$ for the detection of $\mathrm{O_2}$ at $0.76 \ \mathrm{\mu m}$ in the reflection spectrum. We note that $\mathrm{O_3}$ and $\mathrm{O_2}$ are considered to be the most important potential biomarkers for Earth-like planets.

\begin{figure}
  \centering
  \resizebox{\hsize}{!}{\includegraphics{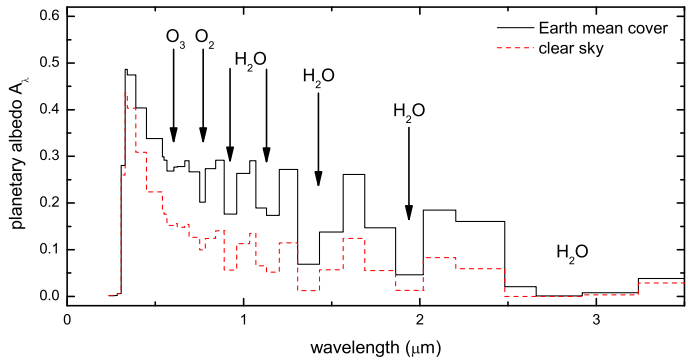}}
  \caption{Planetary albedos at the top of the atmosphere of the Earth model from near UV to IR wavelengths. The dashed line corresponds to the clear sky case, and the solid line indicates the results for a mean Earth cloud cover ($39.5\%$ low-level clouds, $15\%$ high-level clouds, $7\%$ overlap of both cloud layers).}
  \label{earth_ref_spectrum_ratio}
\end{figure}

Since reflection spectra depend on the incident radiation, the wavelength dependent planetary albedos are often analysed in this wavelength region. They are defined by
\begin{equation}
  A_\lambda = \frac{F_{\mathrm{p},\lambda}}{F_{\mathrm{s},\lambda}} \ ,
\end{equation}
where $F_{\mathrm{p},\lambda}$ is the outgoing spectral flux of the planetary atmosphere and $F_{\mathrm{s},\lambda}$ is the stellar spectral flux incident at the top of the atmosphere (see Fig. \ref{stellar_spectra}). For Earth, these albedos are usually obtained from earthshine observations, i.e. by measuring the spectra of the Earth reflected by the lunar surface. The albedos provide information on the atmospheric composition excluding the direct influence of the spectral energy distribution of the incident solar radiation. The planetary albedos related to the reflection spectra of Fig. \ref{earth_ref_spectrum} are shown in Fig. \ref{earth_ref_spectrum_ratio}. The albedo is increased by nearly a factor of three in the VIS caused by clouds. Water droplets and ice cloud particles have an almost constant single scattering albedo of 0.8 throughout the whole VIS-NIR wavelength range (see Fig. 1 in Paper I) and are, therefore, highly reflective. In addition, with an optical depth (at $0.6 \ \mathrm{\mu m}$) of 4.7 and 2.2, respectively, their effect on reflected incident solar radiation is huge. The (mean) surface albedo of Earth is about $\sim0.1$, while the (temporal) average planetary Bond albedo is about $\sim0.3$ (see e.g. \citet{Trenberth2009}). Without clouds the planetary surface would be the major cause of the scattering of incident radiation in the visible wavelength range (neglecting the Rayleigh scattering), hence the planetary albedo would be more or less equal to the surface albedo if no clouds were present. The difference of a factor of three between this clear-sky albedo and the measured mean Earth albedo results from the increased reflection of incident solar radiation by clouds. Thus, the mean Earth cloud cover yields an increase of nearly $300\%$ in the planetary albedo. With $100\%$ low-level clouds, this increase would be even larger as demonstrated using observations and synthetic spectra by \citet{Tinetti2006a}. In contrast to reflection spectra, the absorption bands of $\mathrm{O_2}$ and $\mathrm{H_2 O}$ can be more easily identified in the calculated albedos. This is especially true for the broad Chappuis band of $\mathrm{O_3}$, which is not visible in the low-resolution reflection spectra. With earthshine observations, \citet{Turnbull2006ApJ} were able to detect these $\mathrm{H_2 O}$ bands, as well as the absorption bands of $\mathrm{O_3}$ and $\mathrm{O_2}$.

As already discussed in Paper II, clouds lead to an overall decrease in the outgoing IR flux and a dampening of spectral absorption features in the thermal emission spectrum. In particular, the $\mathrm{O_3}$ band is strongly affected, while other molecular bands, such as $\mathrm{CO_2}$, show only minor changes. This is a direct consequence of the IR radiation being trapped in the lower atmosphere by the high-level clouds, in combination with the lower surface temperatures caused by the albedo effect of the low-level clouds (see Paper II for a detailed discussion on these effects). At shorter wavelengths, clouds have the opposite effect as clearly shown in Fig. \ref{earth_ref_spectrum}. They increase the planetary albedo and, therefore, the amount of back-scattered solar radiation leading to deeper molecular absorption bands, as can be directly inferred from Figs. \ref{earth_ref_spectrum} and \ref{earth_ref_spectrum_ratio} in case of the $\mathrm{O_2}$ A-band. The corresponding radiation flux profiles illustrating in detail these effects are shown in Fig. 5 of Paper I.

The impact of clouds on the reflection spectra and spectral albedos of Earth was previously studied with observations and synthetic spectra by e.g. \citet{Tinetti2006a,Tinetti2006b}, who used a line-by-line radiative transfer code to calculate the synthetic spectra. They found an increase in the spectral albedo in the visible wavelength range of $500\%$ at full cloud coverage (e.g. increase in the clear-sky albedo from 0.06 to $\sim$0.3 at full cloud cover over the Pacific Ocean). Their observations and the simulated spectra confirmed that clouds lead to an overall increase in back-scattered radiation in the visible and NIR wavelength ranges. This also resulted in considerably deeper absorption bands, such as $\mathrm{O_2}$, especially in the presence of low-level clouds. \citet{Kaltenegger2007ApJ} studied the spectrum of Earth at different epochs. In their VIS-NIR mid-resolution spectra ($R=70$) of the epoch corresponding to present Earth conditions, the major observable molecules are $\mathrm{H_2 O}$, $\mathrm{O_2}$, and $\mathrm{O_3}$, whereas for $\mathrm{CH_4}$ and $\mathrm{CO_2}$ no significant absorption features were detected which agrees with our findings (see Figs. \ref{earth_ref_spectrum} and \ref{earth_ref_spectrum_ratio}).

In contrast to the previous line-by-line radiative transfer studies yielding (only) high-resolution spectra, we directly take the impact of clouds into account in the atmospheric model. Consequently, in this study the presence of clouds in the atmospheres of Earth-like extrasolar planets is not only affecting the resulting planetary spectra but also the calculated atmospheric structures.

The two-stream radiative transfer of the plane-parallel atmospheric model used in our study directly calculates disk-averaged spectra. It is therefore unable to take into account the different observational setups (e.g. satellite viewing angle, partial illuminated Earth etc.) that would be necessary for a direct comparison with the Earth observations of \citet{Tinetti2006a}. A realistic quantitative comparison with observed Earth spectra would require a more advanced radiative transfer model. However, our low-resolution spectra and spectral albedos show a good qualitative agreement with these observations in terms of detectable molecules, viz. the strong absorption bands of $\mathrm{H_2 O}$ in the VIS, the Fraunhofer A-band of $\mathrm{O_2}$, and the Chappuis band of $\mathrm{O_3}$. Owing to the high resolution of their spectra, \citet{Tinetti2006a} were also able to identify the weak Fraunhofer B-band of $\mathrm{O_2}$, which, however, is too narrow and too weak to be observable at low resolution. Our results also agree with the principle radiative effects of clouds reported by \citet{Tinetti2006a,Tinetti2006b} and \citet{Kaltenegger2007ApJ}, namely that they cause an increase in scattered radiation and depths of absorption bands. Our simplified radiative transfer model is therefore suitable for studying the basic effects of the considered two cloud types on the reflection spectra of Earth-like extrasolar planets.

The present study is focused on the effect of water and ice clouds in Earth-like planetary atmospheres. Given a present Earth-like chemical composition of the atmosphere, only the molecules $\mathrm{O_2}$, $\mathrm{H_2 O}$, and in principle also $\mathrm{O_3}$ have detectable features in the NUV-VIS wavelength range at low resolution. However, as pointed out by \citet{Kaltenegger2009} $\mathrm{CO_2}$ or $\mathrm{CH_4}$ should also be visible at low resolution if their abundances in the atmosphere were increased\footnote{\citet{Turnbull2006ApJ} were able to identify $\mathrm{CO_2}$ and $\mathrm{CH_4}$ absorption features in their high-resolution spectrum obtained from earthshine observations which, are, however, invisible at low resolution.}. During an earlier epoch of the Earth, $\mathrm{CH_4}$ might have been visible (see \citet{Kaltenegger2007ApJ}), whereas $\mathrm{CO_2}$ could be observed in the NIR of the reflection spectrum of a $\mathrm{CO_2}$ dominated atmosphere such as Venus \citep{Vasquez2010ASPC}.
\begin{figure*}
  \centering
  \resizebox{\hsize}{!}{\includegraphics{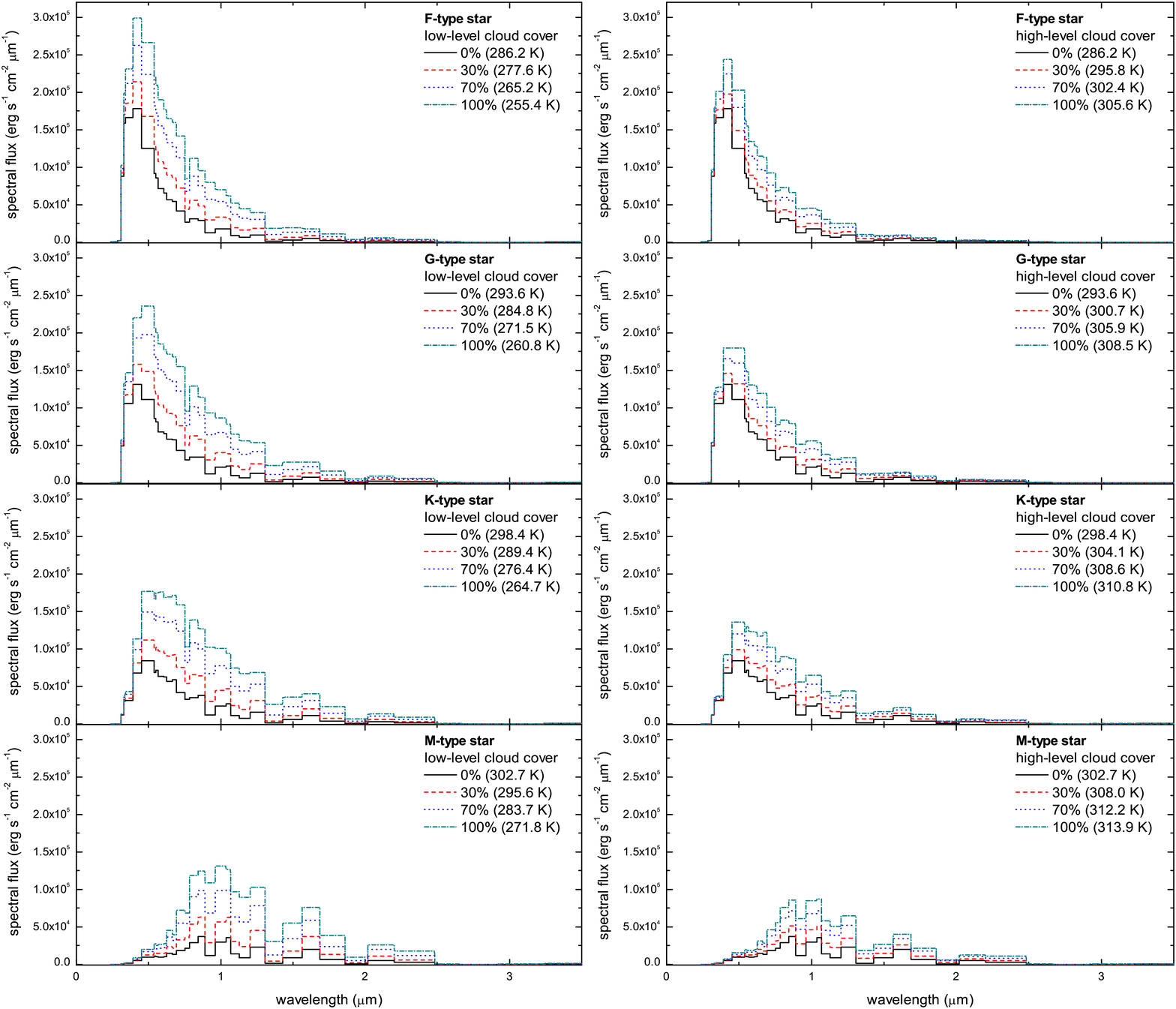}}
  \caption{Reflection spectra influenced by single low-level clouds (\textit{left diagram}) and high-level clouds (\textit{right diagram}). The resulting spectra are shown for each stellar type and four different coverages of the respective cloud type (solid lines: $0\%$; dashed lines: $30\%$; dotted lines: $70\%$; dashed-dotted lines: $100\%$). The planetary surface temperatures resulting from these cloud covers are given in parenthesis.}
  \label{spectra_single_layer}
\end{figure*}

The effect of other cloud types (such as $\mathrm{CO_2}$ ice clouds) depend in general on their optical properties (scattering phase function, scattering and absorption coefficients, and optical depths), which can be very different from those of the water and ice clouds considered here. These properties, though, are complicated functions of particle size distribution, refractive indices, or particle shape, which in turn depend strongly on the atmospheric conditions. The effects of clouds on a reflection spectrum are also largely determined by the cloud altitude within the atmosphere. Only spectral features originating from molecules above the cloud layer(s) can be enhanced in the reflection spectra by the increase in back-scattered radiation caused by cloud particles.

\section{Reflection spectra of cloudy Earth-like extrasolar planetary atmospheres}
\label{sec_spectra}

We now study the effects of single and multi-layered clouds on the reflection spectra and spectral albedos of Earth-like extrasolar planets orbiting different types of central stars. To illustrate the basic effects of the two different cloud types on the reflection spectra and planetary albedos, we first present calculations using only single cloud layers. We then discuss the corresponding effects of multi-layered clouds yielding mean Earth surface temperatures of $288 \ \mathrm{K}$.

\subsection{Single high-level ice and low-level water droplet clouds}
\label{sub_basic_effects}
\begin{figure*}
  \centering
  \resizebox{\hsize}{!}{\includegraphics{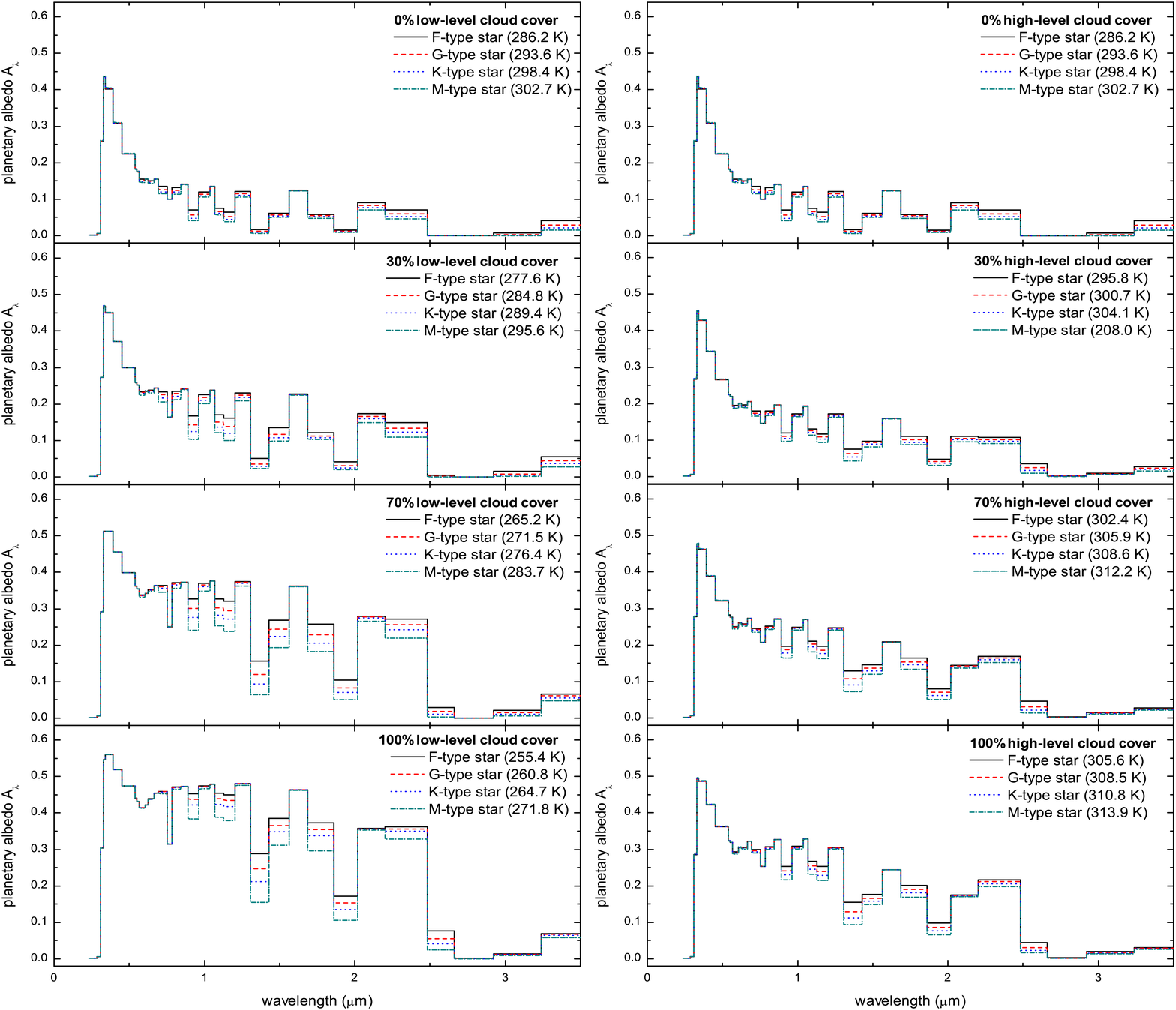}}
  \caption{Wavelength dependent planetary albedo $A_\lambda$ influenced by single low-level clouds (\textit{left diagram}) and high-level clouds (\textit{right diagram}). The albedos are shown for each stellar type and four different coverages of the respective cloud type (solid lines: $0\%$; dashed lines: $30\%$; dotted lines: $70\%$; dashed-dotted lines: $100\%$). The planetary surface temperatures resulting from these cloud covers are given in parenthesis.}
  \label{ratio_single_layer}
\end{figure*}
For each stellar type we calculated low-resolution reflection spectra of Earth-like planets for single low-level water and high-level ice clouds. In Fig. \ref{spectra_single_layer} we depict the resulting spectra for three representative cloud coverages (30\%, 70\%, and 100\%) and the respective clear sky cases. This shows that clouds have a strong impact on the reflection spectra by increasing the amount of reflected incident stellar radiation due to the scattering of cloud particles and leading to deeper molecular absorption bands, such as $\mathrm O_2$ and $\mathrm H_2 \mathrm O$. 
The reflection spectra depend directly on the spectral energy distribution of the incident stellar radiation. The incident stellar spectra, therefore, determine the spectral signatures that can be found in the reflection spectra. 

In case of the M-type star, for example, only the $\mathrm H_2 \mathrm O$ absorption features ($\sim 0.95 \ \mathrm{\mu m}$, $\sim 1.1 \ \mathrm{\mu m}$, $\sim 1.4 \ \mathrm{\mu m}$, and $\sim 2 \ \mathrm{\mu m}$) can be seen at low resolution. In contrast, these bands are almost invisible for the F-type star spectrum because this stellar type emits very little in that wavelength region (see Fig. \ref{stellar_spectra}). The K-type star is particularly interesting because of its spectral maximum near the Chappuis band of $\mathrm O_3$ at $9.6 \ \mathrm{\mu m}$. Therefore, this is the only case where ozone can be directly detected in the low-resolution reflection spectra for high cloud covers. For large cloud coverages the $\mathrm O_2$ signature at $0.76 \ \mathrm{\mu m}$ can also be identified in the spectra for all types of central stars with the exception of the M-star case. 

In general, the low-level water clouds have a larger overall effect on the reflection spectra than the high-level ice clouds\footnote{Note the same axis scaling in Fig. \ref{spectra_single_layer}.} because of their much higher scattering optical depth. Since the optical depths of both clouds is almost constant in the wavelength range of the maxima of the incident stellar radiation\footnote{The detailed optical properties of both cloud types are given in Fig. 1 of Paper I.}  (i.e. $\lambda < 2 \mu \mathrm m$), clouds basically scatter the incident stellar radiation without changing its spectral distribution, in contrast to the effect of Rayleigh scattering. The Rayleigh scattering by molecules results in the well-known blue shift of the reflected light caused by the $\lambda^{-4}$-dependence of the Rayleigh scattering cross-section. For small cloud coverages this effect dominates the reflection spectra.

Figure \ref{ratio_single_layer} also shows that in the M-star case considerably less radiation is back-scattered, even though the overall energy input is the same at the top of the planetary atmospheres for all central stars. This is a direct consequence of the $\lambda^{-4}$-dependence of the Rayleigh scattering. For longer wavelengths, Rayleigh scattering becomes inefficient leading to much less back-scattered incident stellar radiation for the M-type star (maximum of stellar spectrum at about $1 \ \mathrm{\mu m}$, see Fig. \ref{stellar_spectra}) than for instance the F-star case (maximum at ca. $0.4 \ \mathrm{\mu m}$).

The planetary albedos presented in Fig. \ref{ratio_single_layer} have been calculated for the same cloud cover scenarios as in Fig. \ref{spectra_single_layer}. For presentational reasons, the results for all four stellar types are plotted here in a single diagram for each cloud coverage. 
For a given cloud cover, the planetary albedos are obviously almost independent of the spectral energy distribution of the incident stellar radiation in contrast to the reflection spectra (cf. Fig. \ref{spectra_single_layer}). They depend mainly on only the atmospheric chemical composition and the optical properties of the cloud particles. The minor differences in the depths of the water bands are caused by the different temperature profiles and surface temperatures (see Paper II for the corresponding temperature profiles), which determine the water content in the convective region of the lower atmosphere by the relative humidity distribution according to \citet{Manabe67}. 

The planetary albedos are increased by a factor of up to eight in the near IR and up to five in the visible wavelength region relative to the clear sky cases. With a factor of about 1.2, this increase is much smaller in the near UV. This is directly correlated to both the optical properties of the cloud particles, which show a clear decline in the optical depth, and to the more dominating Rayleigh scattering in the near UV wavelength region.

In all clear sky cases, the pronounced $\mathrm H_2 \mathrm O$ bands are almost the only ones visible. With increasing cloud coverage of either cloud type, the depths of the molecular absorption bands are strongly enhanced for all types of central stars. This includes in particular the absorption bands of the biomarkers $\mathrm O_2$ at $0.76 \ \mathrm{\mu m}$ and the Chappuis band of ozone ($0.55-0.70 \ \mathrm{\mu m}$). This broad absorption band of $\mathrm O_3$ becomes noticeable for cloud coverages larger than 30\% but is indistinguishable in the low-resolution reflection spectra with the exception of the K-type star. 

As already discussed, low-level clouds have a larger effect on the planetary albedo than high-level ones because of their higher optical depth, as can clearly be inferred from Fig. \ref{ratio_single_layer}.

In the following subsection, we extend the study to the combined effects of both cloud layers. We focus, in particular, on the subset of the model scenarios that can reproduce mean Earth surface conditions (i.e. surface temperatures of $288 \ \mathrm{K}$). As discussed in Paper I, several combinations of cloud covers of high and low-level clouds can yield this mean Earth surface temperature because of the combination of the differing radiative effects of the cloud layers. 
\begin{figure}
  \centering
  \resizebox{\hsize}{!}{\includegraphics{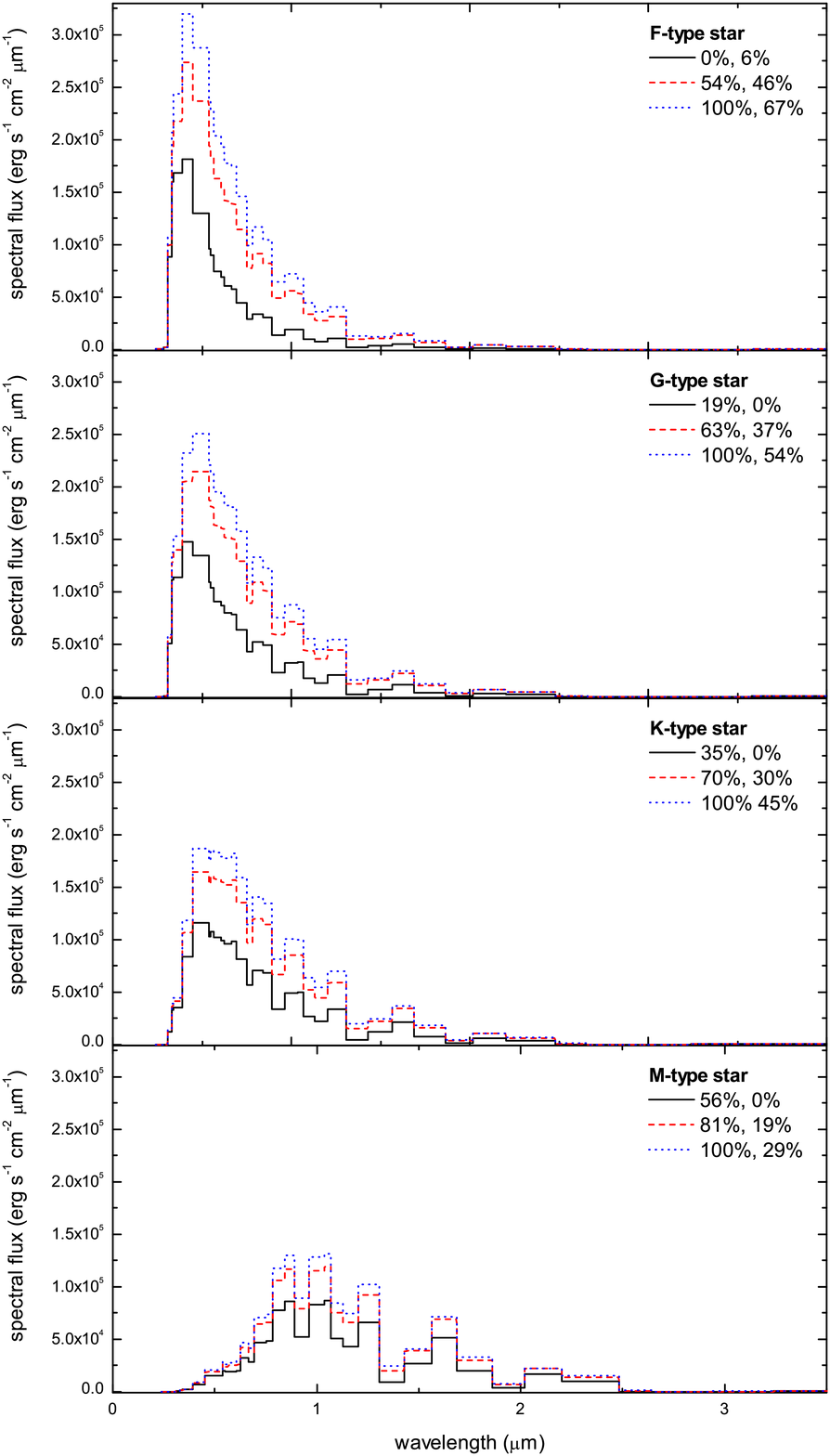}}
  \caption{Reflection spectra influenced by multi-layered clouds with three different cloud-cover combinations (low-level cloud, high-level cloud), yielding surface temperatures of 288 K in all cases.}
  \label{288_spectra}
\end{figure}

For each stellar type, we calculated the spectra for three representative cloud cover scenarios. The cloud cover combinations were chosen to ensure that the possible minimum and maximum cloud coverages are considered for which this particular surface temperature can be obtained. In addition, one intermediate coverage combination has been chosen for each star case. The respective subset of models for which the corresponding spectra are presented in this section is identical to the one used in Paper II (see \citet{Kitzmann2011A&A531} for details). Even though in all of these cases the surface temperatures are identical ($288 \ \mathrm{K}$), the different amounts of cloud coverages of the two cloud types and the differences in the spectral energy distribution of the incident stellar radiation lead to interesting changes in the planetary reflection spectra. In Fig. \ref{288_spectra}, the resulting reflection spectra are summarised. The corresponding planetary albedos and the contrast values between the emissions of the central stars and the planetary reflection spectra, given by the expression
\begin{equation}
  C_\lambda = A_\lambda \left( \frac{R_\mathrm{p}}{d} \right)^2
\end{equation}
where the radius of the planet is $R_\mathrm{p}$ and the distance of the planet to its host star is $d$ (see Sec. \ref{sec_model}), are shown in Fig. \ref{288_ratio}.

\subsection{Effects of multi-layered clouds}
\label{sub_comb_effects}

As discussed in Sect. \ref{sub_basic_effects}, the principle effects of clouds on the reflection spectra are the increases in both the reflected light and the related depths of absorption bands. However, owing to the different cloud covers needed to obtain mean Earth surface temperature conditions the strength of these effects is different for each central star. 

For example, the reflected radiation flux of the F-type star varies by about a factor of three between the minimum and maximum cloud coverage. However, the reflection of light from the planet around the M-type star varies only slightly between its minimum and maximum cloud covers. This is caused by the different cloud cover combinations required for the F star case and the M-type star to reproduce the global mean Earth surface temperature. For the M-type star, a large minimum coverage of low-level clouds is needed to cool the surface, whereas the required minimum high-level cloud cover is quite small in the F star case. 

\begin{figure}
  \centering
  \resizebox{\hsize}{!}{\includegraphics{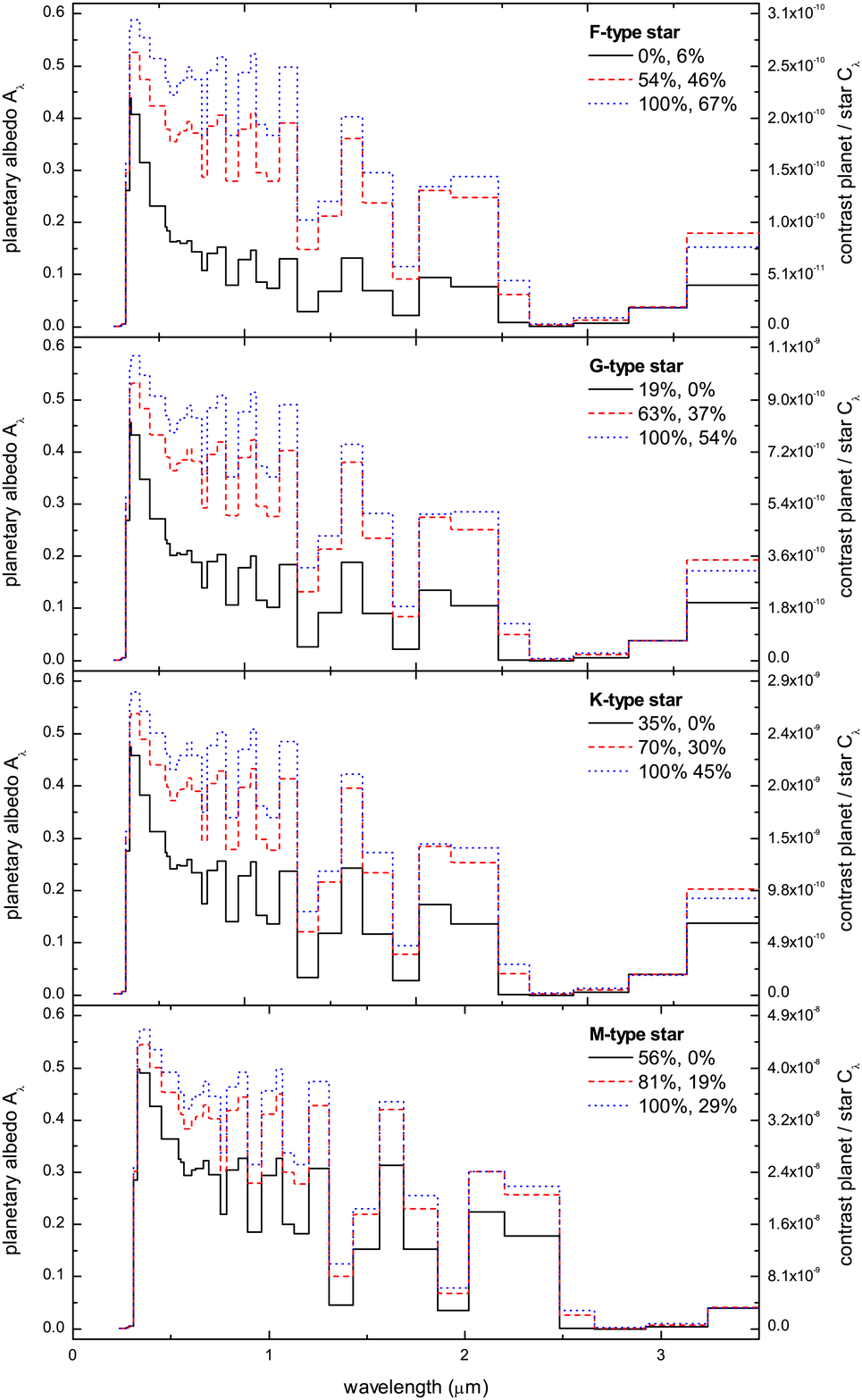}}
  \caption{Planetary albedos A$_\lambda$ (left hand side axis) influenced by multi-layered clouds with three different cloud cover combinations (low-level cloud, high-level cloud) yielding surface temperatures of 288 K in all cases. The axis on the right hand side denotes the contrast C$_\lambda$ between the planetary reflections and the corresponding central star emissions.}
  \label{288_ratio}
\end{figure}

As already mentioned, a wavelength-dependent albedo allows the detection of molecular absorption bands almost independently of the spectral energy distribution of the incident stellar radiation. The deep and broad bands of $\mathrm H_2 \mathrm O$ are again visible in each case (Fig. \ref{288_ratio}). However, the $\mathrm O_3$ band is not directly detectable in case of the smallest amount of clouds for the F (9\% high-level cloud cover) and G (19\% low-level cloud cover) stars. Owing to the large amount of low-level clouds needed to reproduce the mean Earth surface temperature observed, the absorption bands of the biomarkers $\mathrm O_3$ and $\mathrm O_2$ can be seen for the M-type and K-type stars for every considered combination of cloud covers.

The second axis on the right hand side of Fig. \ref{288_ratio} shows in addition the contrast $C_\lambda$ between the reflected planetary and incident stellar radiation fluxes. The ratio between the radiation fluxes of the planet and the central star is quite small because of the much higher stellar luminosities compared to the planetary reflection spectra. In the clear sky cases, the contrast varies between $10^{-10}$ for the F-type star and $10^{-8}$ for the M-type star because the luminosity of the F-type star is much higher than that of the M-type star. The Earth-like planet around the M star is in addition located much closer to its host star as discussed in Paper I. 

Clouds do increase the contrast by about one order of magnitude owing to the already discussed enhancing scattering effects. However, the resulting higher contrast values ($10^{-9}$ F-type star - $10^{-7}$ M-type star) are still very small. Therefore, it is unlikely that these reflection spectra could be measured by the current generation of telescopes, even for the extremely high contrasts produced by clouds.

\section{Summary}

We have combined a 1D radiative-convective atmospheric model with our previously developed parametrised cloud description (Paper I) to study the influence of low and high-level clouds on the reflection spectra and spectral albedos of Earth-like extrasolar planets orbiting different types of central stars.

In view of habitability NUV-NIR reflection spectra are of particular importance for the (possible) detection of the most important potential biomarker molecules $\mathrm O_2$ and $\mathrm O_3$ (Fraunhofer A-band at $0.76 \ \mathrm{\mu m}$ and the Chappuis band at $0.55-0.70 \ \mathrm{\mu m}$). The detectability of molecular features in this wavelength regime is strongly supported by the presence of clouds. Their high scattering efficiency substantially increases the reflected light and the related depths of the absorption bands, in comparison to the respective clear sky conditions where only the broad bands of $\mathrm H_2 \mathrm O$ are visible. Low-level clouds have thereby a larger impact on the spectra than the high-level clouds because of their much higher scattering optical depth. 

However, the visibility of chemical species in low-resolution reflection spectra also depends strongly on the spectral spectral distribution of the incident stellar radiation, i.e. on the type of central star. The reflection spectrum of an Earth-like planet around the M-type, for example, displays no $\mathrm O_3$ or $\mathrm O_2$ feature owing to the lack of incident stellar radiation in the respective wavelength range. Only in case of the K-type star can the $\mathrm O_3$ Chappuis-band be seen directly in the low-resolution reflection spectra at high cloud coverages.

In contrast to the case of reflection spectra, the spectral planetary albedos enable molecular features to be detected, excluding the direct influence of the spectral distribution of the stellar radiation. In particular, we investigated the impact of multi-layered clouds for cloud cover combinations yielding global mean Earth surface temperatures of $288 \ \mathrm{K}$ for each stellar type.
Owing to the large amount of low-level clouds necessary for this surface temperature, the absorption bands of $\mathrm O_3$ and $\mathrm O_2$ are noticeable in the low-resolution spectral albedos for K and M-type star cases, whereas $\mathrm O_3$ is not detectable at low cloud coverages yielding mean Earth surface temperatures for F and G-type star spectra.

Owing to the much higher stellar luminosities compared to the reflected light from the planet, the ratios between the radiation fluxes of the planets to their respective central stars are quite low at clear sky conditions (from $10^{-10}$ (F-star) up to $10^{-8}$ (M-star)). Clouds increase this contrasts by about one order of magnitude but the resulting contrast ratios are still too small to be observable with the current generation of telescopes. However, investigating the polarised radiation scattered by clouds in comparison to the unpolarised stellar radiation may provide an opportunity to characterise cloudy exoplanetary atmospheres in the not so far future \citep[see][and references therein for more details]{Stam2008A&A}.

\begin{acknowledgements}
  This work has been partly supported by the research alliance \textit{Planetary Evolution and Life} of the Helmholtz Association (HGF).
\end{acknowledgements}

\bibliographystyle{aa} 
\bibliography{references}

\end{document}